\documentclass[11pt]{revtex4-1}

\usepackage{braket}
\usepackage{soul}
\newcommand\tr{\operatorname{tr}}
\usepackage{setspace}
\usepackage{amsmath,amssymb,graphicx}

\begin{document}
\title{Structure and efficiency in bacterial photosynthetic light-harvesting}
\author{Susannah Bourne Worster$^\text a$}
\author{Clement Stross$^\text{a,b}$} 
\author{Felix M. W. C. Vaughan$^\text{a,b,c}$}
\author{Noah Linden$^\text b$}
\author{Frederick R. Manby$^\text{a,1}$}

\affiliation{a: Centre for Computational Chemistry, School of Chemistry, University of Bristol, Bristol BS8 1TS, UK}
\affiliation{b: School of Mathematics, University of Bristol, Bristol BS8 1TW, UK}
\affiliation{c: Bristol Centre for Complexity Sciences, University of Bristol, Bristol, BS2 8BB, UK}

\email{To whom correspondence should be addressed. E-mail: fred.manby@bristol.ac.uk}

\begin{abstract}
Photosynthetic organisms use networks of chromophores to absorb sunlight and deliver the energy to reaction centres, where charge separation triggers a cascade of chemical steps to store the energy.  We present a detailed model of the light-harvesting complexes in purple bacteria, including explicit interaction with sunlight; energy loss through radiative and non-radiative processes; and dephasing and thermalizing effects of coupling to a vibrational bath. An important feature of the model is that we capture the effect of slow vibrational modes by introducing time-dependent disorder. 
Our model describes the experimentally observed high efficiency of light harvesting, despite the absence of long-range quantum coherence. The one-exciton part of the quantum state fluctuates due to slow vibrational changes, but remains highly mixed at all times.
This lack of long-range coherence suggests a relatively minor role for structure in determining the efficiency of bacterial light harvesting. To investigate this we built hypothetical models with randomly arranged chromophores, but still observed high efficiency when typical nearest-neighbour distances are comparable with those found in nature.  
This helps to explain the efficiency of energy transport in organisms whose chromophore networks differ widely in structure, while also suggesting new design criteria for efficient artificial light-harvesting devices.
\end{abstract}

\maketitle

Photosynthesis is the process by which organisms absorb energy in the form of sunlight and convert it into chemical energy, and it is the engine that produces almost all global biomass. Since the molecular architecture needed to carry out photosynthesis is costly to produce and maintain, sunlight is initially captured by an extensive antenna network of  chlorophyll molecules (chromophores), through which absorbed energy must then be transported to specialised reaction centres to begin the photosynthetic reaction cycle. Experiments show that transport through the antenna can be highly efficient: absorption of a sufficiently energetic photon almost always leads to the charge-separation process that is a prerequisite for chemical storage of the energy \cite{Wientjes2013,Rijgersberg1980,Fassioli2013a}.

Two-dimensional spectroscopic investigations present the possibility of a substantial role for quantum coherence in achieving high efficiency \cite{Engel2007,Lee2007}, an idea that was initially supported by a variety of theoretical and computational studies \cite{Calhoun2011,Harel2012,Panitchayangkoon2010,Strumpfer2012a,Collini2010,VanHulst2013a}. More recently this has been contested on the grounds that the oscillatory signature observed in the spectroscopy measurements might have been incorrectly interpreted \cite{Stross2016,Christensson2012,Duan2017,Fassioli2013a}, or might be a result of the (unnatural) laser pulses used to excite the sample \cite{Mancal2010,Brumer2012}. Furthermore, even if the oscillatory pattern truly revealed quantum coherence, it might not necessarily have an important functional role \cite{Kassal2013,Tiersch2012}.  The research nevertheless sparked considerable interest in characterising the quantum state of the antenna over time and identifying criteria for efficiency. This has proved a challenging undertaking due to the complexity of the system and the number of factors influencing its behaviour. In particular, it has been shown that the nature of the light source \cite{Mancal2010,Brumer2012}, the method used to model the vibrational environment \cite{Fassioli2009}, the inclusion (or not) of energetic and spatial disorder \cite{Stross2016,Sarovar2013}, and the consideration of short-time versus steady-state dynamics \cite{Manzano2013,Brumer2018,Kassal2013,Mancal2010b,Brumer2012} all influence the way the system evolves. We aim to bring all these ideas together in a comprehensive and realistic model that includes detailed descriptions of each step of the process; from the continuous absorption of weak, incoherent sunlight to the final consumption of excitons at a reaction centre. We apply this model to a larger system than has yet been considered, including both of the major purple-bacteria light harvesting systems (LHI and LHII) and the branched chromophore chains of the reaction centre (RC).

In addition to a drive to understand the functioning of natural photosynthetic antennae, the suggestion that coherence may play a role in exciton transport inspired several attempts to design artificial systems that exploited the same principles. These systems, both hypothetical and real, require exquisite control over the placement and orientation of pigment chromophores in order to create and preserve the necessary coherence \cite{Sarovar2013,Buckhout-White2014,Hemmig2016,Olejko2017}. This makes them difficult to realise and those that have been created are small and delicately constructed. In the second part of this paper we consider an artificial antenna system of much simpler construction: a box of chromophores placed randomly (as if in solution) around a central reaction centre. Remarkably, the high efficiencies we observe demonstrate that structure (in the sense of a detailed arrangement of chromophores) does not play a significant role in determining the efficiency of an antenna. This could offer a pathway to designing cheaper, more scalable synthetic light harvesting devices.

\section*{Theory}
The flow of energy through a photosynthetic antenna system is governed by a complex interplay of several physical processes: coherent evolution of non-stationary states; relaxation due to coupling with vibrational modes in the environment; absorption and emission of light; non-radiative decay, which results in exciton loss, and, finally, consumption of excitons at the reaction centre, triggering the biochemical cascade that ultimately leads to production of adenosine triphosphate. For reasons justified below, we model the system using a Lindblad master equation with a time-dependent Hamiltonian. Thus the evolution of the system density matrix is described by the Liouville-von Neumann equation
\begin{equation}
    \frac{d\rho(t)}{dt} = -\frac i\hbar\left[\hat{H}(t),\rho(t)\right] + \left(\mathcal{L}_\text{rad} + \mathcal{L}_\text{bath} + \mathcal{L}_\text{nr} + \mathcal{L}_\text{sink}\right)\rho(t).
\end{equation}

Although each chromophore is a large, complex molecule, its participation in exciton transport is dominated by its electronic ground state \(\ket{\psi_0}\) and first excited state \(\ket{\psi_1}\) under biologically relevant conditions. Consequently, we model each chromophore site as a two-level system, with the states of a network of chromophores modelled as Hartree products of the single-site states. We consider only the \(n+1\) states with either zero (\(\ket{0}=\ket{0_1,0_2,\cdots,0_n}\)) or one (\(\ket{i}=\ket{0_1\cdots1_i\cdots0_n}\)) exciton \cite{Tiersch2012}. Rather than focusing on a single component, as has been done previously, we consider transport across the whole antenna using a model system consisting of one LHII and one LHI/RC complex ({\it SI Appendix, Fig. S1}). The system contains a total of $n = 65$ chromophores.

\subsection*{Unitary dynamics}
The unitary dynamics of the exciton are governed by a Hamiltonian of the form
\begin{equation}
\hat{H} = \sum_i{E_i}\ket{i}\!\bra{i} + \sum_{i\ne j}V_{ij}\ket{j}\!\bra{i}\label{eq:ham}
\end{equation}
where \({E_i}\) are the excitation energies at each site and \({V_{ij}}\) is the strength of the interaction between the transition dipole moments \(\boldsymbol\mu_i\) on sites \(i\) and \(j\), calculated using the point dipole approximation. The point dipole approximation overestimates the nearest neighbour coupling in the LHI and RC complexes so these values were corrected to match the results of more detailed calculations ({\it SI Appendix}). Realistic disorder is introduced into the Hamiltonian by randomly drawing both \(E_i\) and \(|\boldsymbol\mu_i|\) from distributions computed using time-dependent density functional theory (TDDFT) on snapshots of a molecular dynamics (MD) simulation \cite{Stross2016}. The unit vectors \(\hat{\boldsymbol\mu}_i\) are taken separately from randomly selected MD snapshots.

\subsection*{Interaction with light}
Different light sources induce very different dynamics  \cite{Brumer2018,Brumer2012,Mancal2010}. Coherent pulses,  localised in time and space, excite coherent superpositions of energy eigenstates, which then evolve coherently. Continuous incoherent illumination excites stationary eigenstates with well-defined energies. To understand the dynamics of photosynthesis it is therefore crucial to use a realistic model of the interaction between the system and incident radiation.

We capture this interaction using a Lindblad operator ($\mathcal L_\text{rad}$) derived from the Born-Markov quantum optical master equation, including terms for both absorption ($\mathcal L_\text{rad}^\text{abs}$) and spontaneous and stimulated emission ($\mathcal L_\text{rad}^\text{emit}$). The rates of each process are parameterised to correctly account for the initial blackbody temperature of the sun, the attenuation of the radiation as it passes through the atmosphere, and the anisotropy of absorption into different eigenstates ({\it SI Appendix}).

At \(~10^{-4}\) ps\(^{-1}\) the rate of absorption is orders of magnitude slower than any of the other processes governing the dynamics of the antenna. As a result, the system remains almost entirely in the zero-exciton ground state for all time. We find that the amount of excited state present at any given time is on the order $p=10^{-9}$.\footnote{A simple kinetics calculation gives the steady-state probability of being in the excited state as \(p=k_\text{abs}/(\Gamma_\text{sink} + \Gamma_\text{nr} + k_\text{spont} + k_\text{stim} + k_\text{abs})\), where \(k_\text{spont}\), \(k_\text{stim}\) and  \(k_\text{abs}\) are, respectively, the rates of spontaneous emission, stimulated emission and absortion. An order of magnitude estimate gives \(p = 10^{-9}\).} To accelerate equilibration to the correct excited-state population, simulations are started from an initial state \(\rho_\text{init} = (1-p)\ket{0}\!\bra{0} + p\ket{\psi}\!\bra{\psi}\) with the excited state manifold in a completely delocalised pure state \(\ket{\psi} = 1/\sqrt{n}\sum_j\ket{j}\). As will be seen, the dynamics soon washes out any specific signatures of this initial state.

\subsection*{Capture of excitons at the reaction centre}
At the reaction centre, excitons are used to trigger a series of electron transfers. As this process is effectively irreversible \cite{Blankenship2011}, we treat the reaction centre as a sink through which excitons are continuously lost. It is described by the Lindblad operator
\begin{equation}
\mathcal{L}_\text{sink}\rho = \Gamma_\text{sink}\sum_{j\in \text{SP}} \left(\ket{0}\!\langle{j}|\rho|{j}\rangle\!\bra{0}-\frac12\{\ket{j}\!\bra{j},\rho\}\right),\label{eq:sinklind}
\end{equation}
where the sum runs over the chromophores in the `special pair', the rate of exciton loss \(\Gamma_\text{sink}\) is taken as the rate of the first step in the electron transfer chain \cite{Greenfield2002,Demmig-Adams2014}, and $\{\cdot,\cdot\}$ denotes anticommutator.

\begin{table}[b]
\centering
\caption{Rate constants and timescales for Lindblad terms}\label{tab:rateconstants}
\begin{tabular}{lccc}
\hline
 & \(\Gamma_\text{sink}\) & \(\Gamma_\text{nr}\) & \(\Gamma_\text{deph}\) \\
\hline
Value / ps\(^{-1}\) & 0.125 & 0.001 & 11  \\
References & \citenum{Greenfield2002,Demmig-Adams2014} & 
\citenum{Palacios2002,Zankel2006,Pandit2011,Monshouwer2002,Bopp1997} &
 \citenum{Harel2012,Fidler2013}\\
 Timescale & 8 ps & 1 ns & 90 fs \\
\hline
\end{tabular}
\end{table}

\subsection*{Coupling to the environment}
The vibrational modes of the light-harvesting complexes and the surrounding medium have a huge impact on how excitons are transported through the antennae. They can act both to reduce the amount of excited state in the system by offering a pathway for non-radiative decay back to the ground state, and to redistribute remaining excitations by acting as a driving force for relaxation within the excited state manifold. Vibrations that are slow on the timescale of dynamics relevant for photosynthetic energy transport disrupt the symmetrical ring structure of the individual light-harvesting complexes, introducing variations in the excitation energies of individual chromophores, and in the interactions between them. We include each of these effects separately in our model.

The symmetry-breaking effects of slow vibrations are captured by introducing static disorder into the Hamiltonian, as described above. The exact form of the disorder will vary slowly with the motion of the vibration over the lifetime of an exciton. We approximate this time-dependence by generating a new, random Hamiltonian at regular intervals of \(0.1\) ps, corresponding approximately to a vibrational frequency below which higher harmonic levels become significantly occupied at 300 K. To our knowledge, this is the first time any attempt has been made to include the dynamical effects of slow vibrations in a master equation description of photosynthetic exciton transfer, although a similar technique has been used successfully to describe the photochemistry of small organic molecules \cite{Schile2019}.

Non-radiative loss of excitons competes with their biochemical utilisation at the reaction centre. We model it using a Lindblad operator \(\mathcal{L}_\text{nr}\) that has an identical form to \(\mathcal{L}_\text{sink}\) (equation \ref{eq:sinklind}), with \(j\) summed over every site in the system and a non-radiative decay constant \(\Gamma_\text{nr}\).

We consider two different models for the Lindblad term \(\mathcal{L}_\text{bath}\) describing the relaxation processes. The first is  a local dephasing model
\begin{equation}
\mathcal{L}_\text{deph}\rho = \Gamma_\text{deph}\sum_j\left(\ket{j}\!\bra{j}\rho\ket{j}\!\bra{j} - \frac12\{\ket{j}\!\bra{j},\rho\}\right),\label{eq:dephlind}
\end{equation}
in which the coherences between the states \(\ket{j}\) localised on each site decay exponentially with a decay constant \(\Gamma_\text{deph}\). Our second model is a global thermalizing model \cite{Ostilli2017a}
\begin{equation}
\mathcal{L}_\text{therm}\rho = \sum_{a,b, a\neq b} \frac{k_{b\rightarrow{}a}}{2\pi}\left(\ket{a}\!\langle{b}|\rho|{b}\rangle\!\bra{a} - \frac12 \{\ket{b}\!\bra{b},\rho\}\right),\label{eq:globlind}
\end{equation}
in which exciton population is transferred between eigenstates \(\ket{b}\) and \(\ket{a}\) at a rate \(k_{b\rightarrow a}\). The local model provides an intuitive picture of the effect of random environmental fluctuations but forces the system to relax to the maximally mixed state \(\rho_\text{mm} = \mathbb{I}/n\) ({\it SI Appendix}). By contrast, the global approach relaxes the system to the correct instantaneous thermal equilibrium state but does not explicitly account for the proximity and strength of coupling between chromophores. In a regime where inter-chromophore couplings are small on the scale of the system-bath coupling, this can result in an unphysical flow of energy between spatially well-separated and uncoupled chromophores \cite{Hofer2017}. 

A further concern is the possible role of non-markovianity in the system-bath coupling \cite{Caruso2010, Vaughan2017}. 
However we find that the global Lindblad approach gives good qualitative agreement with preliminary calculations on small model systems using the exact hierarchical-equations-of-motion (HEOM) approach \cite{FelixThesis}.

The rate constants in equation \ref{eq:globlind} are determined using the Redfield theory expression \cite{May2011}
\begin{align}
k_{a\rightarrow{}b} = & \exp\left(\frac{\omega_{ab}}{k_\text{B}T}\right)k_{b\rightarrow{}a} \notag\\
= & 2\pi\omega_{ab}^2\left((1+n(\omega_{ab}))J(\omega_{ab}) + n(\omega_{ba})J(\omega_{ba})\right),\label{eq:Redfieldrates}
\end{align}
where the relationship between \(k_{a\rightarrow{}b}\) and \(k_{b\rightarrow{}a}\) ensures that detailed balance is obeyed. The angular frequency \(\omega_{ab}\) is that of the energy gap between states \(\ket{a}\) and \(\ket{b}\) and \(n(\omega) = 1/(\exp(\omega/k_\text{B}T)-1)\) is the Bose-Einstein distribution.

The spectral density \(J(\omega)\) describes the coupling of the excitonic system to the vibrational modes of the environment, under the assumption that all modes can be modelled as harmonic oscillators. This assumption breaks down for flat-bottomed, low-frequency modes, whose effect, however, we capture separately by periodically changing the Hamiltonian, as described above. In the absence of a full spectral density for the antenna system, we use the form proposed by Renger and Marcus \cite{Renger2002}:
\begin{equation}
J_\text{RM}(\omega) = 
0.8\frac{\omega^3}{7!\, 2\omega_1^4}
e^{-\sqrt{\omega/\omega_1}} +
0.5\frac{\omega^3}{7!\, 2\omega_2^4}
e^{-\sqrt{\omega/\omega_2}}, 
\label{eq:rengermarcus}
\end{equation}
which is fitted to the experimental spectra of a monomer pigment-protein complex (B777) with characteristic frequencies \(\omega_1 = 0.10483\text{ rad ps}^{-1}\) and \(\omega_2 = 0.364625\text{ rad ps}^{-1}\). Using this spectral density, the global model in equation \ref{eq:globlind} accurately reproduces the experimentally measured relaxation time constant for an LHII ring \cite{Stuart2011, Pullerits1994,Freiberg1998} ({\it SI Appendix, Fig. S2}).

\subsection*{Efficiency}
The efficiency of a photosynthetic apparatus can be defined in a number of different ways, not all of which are equivalent \cite{Manzano2013}. In this work, we will define the efficiency as
\begin{equation}
\mathcal{E}[\rho]=\frac{\braket{0|\mathcal{L}_\text{sink}\rho|0}}{\braket{0|(\mathcal{L}_\text{sink} + \mathcal{L}_\text{nr} + \mathcal{L}_\text{rad}^\text{emit})\rho|0}}.
\end{equation}
For each unit of energy that is absorbed by the antenna, the efficiency tells us what fraction is used productively by the sink. This measure reflects how well the system is adapted to transport and use the absorbed energy. Note that it does not tell us what fraction of incident light reaches the reaction centre as this would require additional consideration of how efficiently the system absorbs light. While this is also an important question, we will not consider it here.

\section*{Results}

\subsection*{Bacterial antenna system}

\begin{figure}
\centering
\includegraphics[width=.8\linewidth]{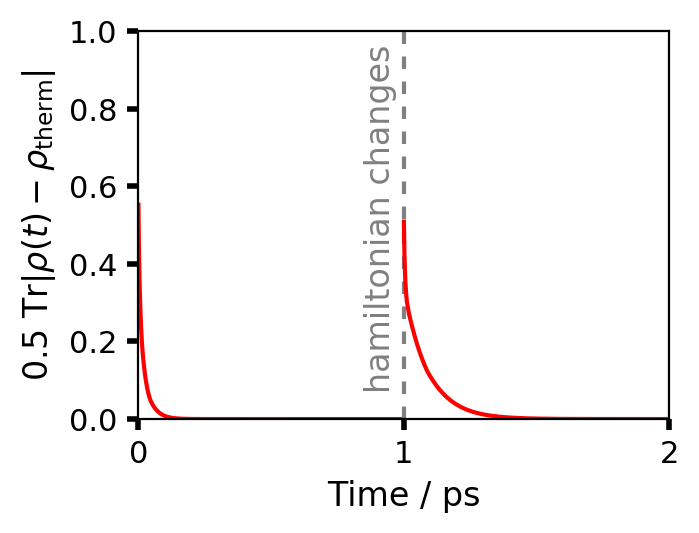}
\caption{Trace distance between the time-evolving state and the thermal state \(\rho_\text{therm} = \sum_\alpha\exp(E_\alpha/k_\text{B}T)/Z\) of the current Hamiltonian as a function of time for the LHI-LHII model system. The Hamiltonian changes at 1 ps, as does the corresponding \(\rho_\text{therm}\). It takes several hundred femtoseconds for the system to requilibrate, under the influence of the Hamiltonian and the thermalizing Lindblad operator \(\mathcal{L}_\text{therm}\). }
\label{fig:relaxationrates}
\end{figure}

Under the influence of \(\mathcal{L}_\text{bath}\), the excited-state manifold relaxes rapidly towards a steady-state. When \(\mathcal{L}_\text{bath} = \mathcal{L}_\text{deph}\), the dynamics for every realization of \(\hat{H}\) equilibrates to \(\rho_\text{mm}\). 

\begin{figure}[t]
\centering
\includegraphics[height=10.4cm]{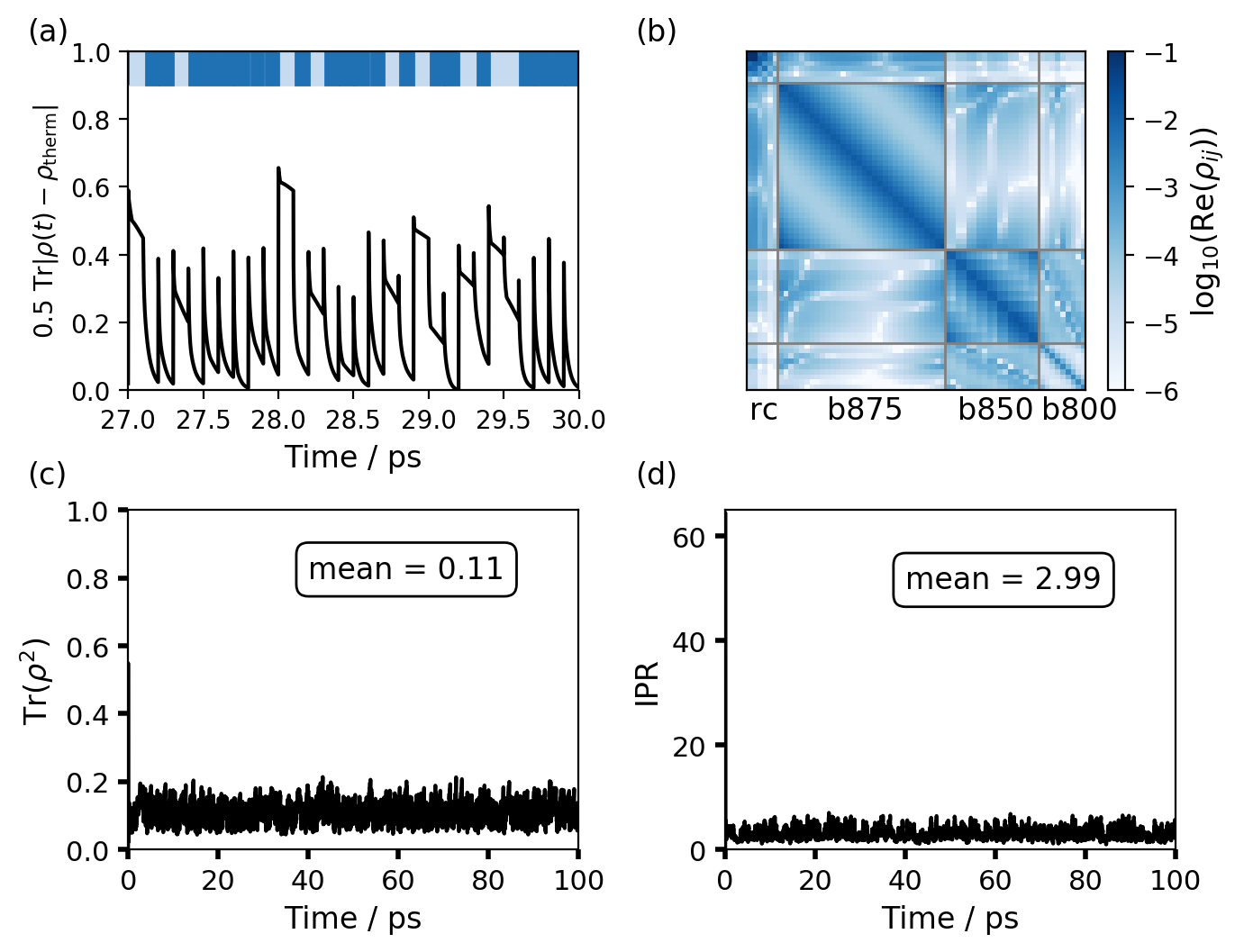}
\caption{Evolution of the state of the exciton manifold in the full LHI-LHII model system. The Hamiltonian changes every 0.1 ps. Dephasing is implemented using the global model \(\mathcal{L}_\text{therm}\) (an equivalent figure for the local model can be found in {\it SI Appendix, Fig. S5}). Rates of other processes are given in table \ref{tab:rateconstants}.  (a) Trace distance of the instantaneous evolving state \(\rho(t)\) from the current thermal state \(\rho_\text{therm}\), which changes with the Hamiltonian (only a short time sample is shown so that the details of the evolution are more clearly visible). The coloured bar along the top illustrates the amount of time spent close to a thermal state, with dark blue marking stretches of time where the trace distance is \(\leq 0.2\), and light blue where the trace distance is \(> 0.2\). (b) Plot of \(\log_{10}|\operatorname{Re}(\rho_\text{av})|\). The renormalized average density matrix \(\rho_\text{av}\) is calculated from 5 ps onward, to avoid any influence from the specific choice of initial state. (c) Degree of purity quantified by \(\tr\rho(t)^2\). (d) The inverse participation ratio of \(\rho(t)\). For (c) -- (d), the mean value of each quantity starting from 5 ps is displayed on the graph.}\label{fig:propertiesaveragestate}
\end{figure}

When \(\mathcal{L}_\text{bath} = \mathcal{L}_\text{therm}\) the steady-state is the thermal state of the current Hamiltonian, in which eigenstates are populated according to the Boltzmann distribution. The thermal state changes when the Hamiltonian changes, with an average trace distance of 0.4 between instantaneous thermal states. Following a change in the Hamiltonian, the exciton manifold reequilibrates to the new thermal state within a few hundreds of femtoseconds (see figure \ref{fig:relaxationrates}), in keeping with experimentally measured equilibration times in individual LHCs \cite{Stuart2011}.  The Hamiltonian changes on a similar timescale (every 100 fs), such that the quantum state of the antenna is constantly evolving (see figure \ref{fig:propertiesaveragestate}a). However, since individual, instantaneous thermal states are not too dissimilar, the system spends much of its time in states that are close to thermal and the  time-averaged state \(\rho_\text{av}\), shown in figure \ref{fig:propertiesaveragestate}b, is close (trace distance \(\approx\) 0.15) to the average of the thermal states for all realisations of the Hamiltonian \(\braket{\rho_\text{therm}}_\text{av}\). The longer the period between changes of the Hamiltonian, the more closely \(\rho_\text{av}\) resembles \(\braket{\rho_\text{therm}}_\text{av}\).

Thermal states are, by nature, mixed states. The extent to which the state of a system is mixed is quantified by $\tr\rho^2$, which is 1 for a pure state and \(1/N\) for the maximally mixed state. It may be shown analytically ({\it SI Appendix}) that the full average over the thermal states of all possible Hamiltonians is exactly the maximally mixed state. Correspondingly, the values of $\tr\rho^2$, given in table \ref{tab:subsystemstates}, for the average states of both the full system and each component subsystem are small and close to their values for the maximally mixed state. On average then, the system is not only in a highly mixed state but in one that equally favours equivalent sites within a ring.

\begin{table}
\centering
\caption{Properties of subsystem states calculated for the renormalized substates of the average state}\label{tab:subsystemstates}
\begin{tabular}{lrrrr}
        & LHI-LHII & B875 & B850 & B800 \\
\hline
$\tr\rho^2$ & 0.083 & 0.095 & 0.141 & 0.114 \\
Min. value of $\tr\rho^2$ & 0.015 & 0.03 & 0.06 & 0.11 \\
IPR & 3.60 & 8.86 & 6.73 & 1.57\\
Max. value of IPR & 65 & 32 & 18 & 9\\
\hline
\end{tabular}
\end{table}

The average state, furthermore, contains very little coherence, as evidenced by small values of the inverse participation ration (IPR). The IPR, \cite{Meier2002}
\begin{equation}
\text{IPR}=\frac{\left[\sum_{jk}|\rho_{jk}|\right]^2}{n_\text{site}\sum_{jk}|\rho_{jk}|^2},
\end{equation}
provides a measure of how coherently delocalised a state is. It is bounded between 1 (incoherent and fully localised) and \(n_\text{site}\) (fully delocalised coherent pure state). In B800 particularly, there is essentially no coherence (IPR = 1.57), reflecting the fact that the chromophores are well-spaced and weakly coupled. Stronger coupling in the B850 and B875 rings introduces coherences between neighbouring sites, raising the value of the IPR. However, when \(\rho_\text{B850}\) and \(\rho_\text{B875}\) are expressed in the reduced basis of the BChl dimer subsystem, to more accurately reflect the structure of these rings, they are also found to be essentially maximally mixed ({\it SI Appendix, Fig. S3}).

Overall, the excitonic subsystem is, on average, in a highly mixed, incoherent state that nevertheless places population equally on all sites. Instantaneously, the system is never in its average state but rather fluctuates continuously between instantaneous thermal states. However, since the thermal states all have similar forms (though favouring different chromophore sites), the properties of the system do not fluctuate a great deal. Plots of $\tr\rho(t)^2$ and IPR in figure \ref{fig:propertiesaveragestate} (panels c and d) indicate that, at all times, the state of the system remains highly mixed with no long-range coherence.

\begin{figure}
\centering
\includegraphics[width=.8\linewidth]{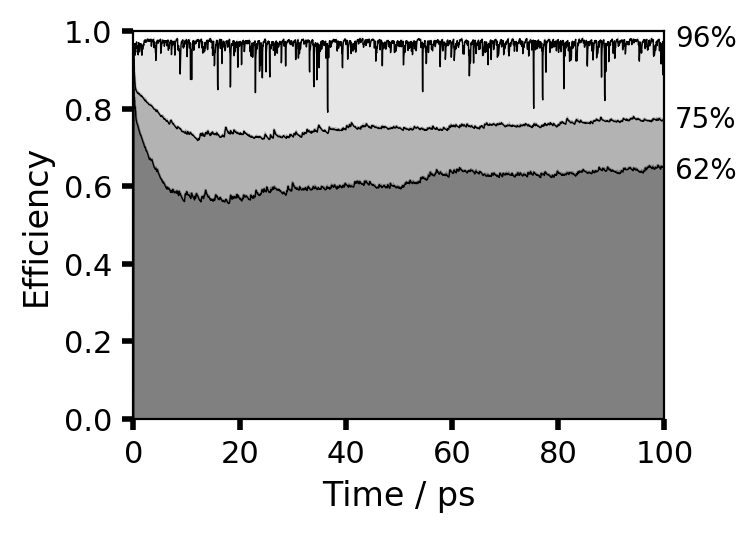}
\caption{Efficiency of the LHI-LHII antenna over the same period of evolution plotted in figure \ref{fig:propertiesaveragestate}. From top to bottom, the traces correspond to efficiency of the full antenna (one LHI and one LHII) with \(\mathcal{L}_\text{bath}=\mathcal{L}_\text{therm}\); efficiency of the LHI/RC complex with \(\mathcal{L}_\text{bath}=\mathcal{L}_\text{deph}\), and efficiency of the full antenna with \(\mathcal{L}_\text{bath}=\mathcal{L}_\text{deph}\). The mean efficiency, displayed to the right hand side of each trace, is calculated from 5 ps.}
\label{fig:antennaefficiency}
\end{figure}

The efficiency of the antenna, plotted in figure \ref{fig:antennaefficiency}, remains fairly constant over time, reflecting the consistent form of the quantum state throughout the evolution. The efficiencies obtained with the two variants of \(\mathcal{L}_\text{bath}\) span the experimentally measured efficiencies for photosystem II in higher plants (84\%--91\%) \cite{Fassioli2013a,Wientjes2013}. 

Using \(\mathcal{L}_\text{bath}=\mathcal{L}_\text{therm}\), the time-averaged efficiency \(\braket{\mathcal{E}[\rho(t)]}_\text{av}\) is 96\%. This is likely an overestimate, since the global model does not account for the weaker coupling between B875 and the RC, which significantly retards the rate of exciton transfer between them \cite{Blankenship2002}. Hence, this model tends to overestimate the population of the RC ({\it SI Appendix, Fig. S4}), on which \(\mathcal{E}\) is heavily dependent ({\it SI Appendix}).

By contrast, using \(\mathcal{L}_\text{bath}=\mathcal{L}_\text{deph}\) underestimates the efficiency by failing to capture the energy gradient between LHII and LHI and consequently underestimating the population of the B875 ring and the rate of exciton transfer onto the RC. Using the local dephasing model \(\braket{\mathcal{E}[\rho(t)]}_\text{av}\) is 62\%. This value is highly sensitive to the number of antenna chromophores, since exciton density is distributed equally between all chromophores under the action of \(\mathcal{L}_\text{deph}\). Therefore, increasing the number of chromophores reduces the population on any single site. For example, if only LHI is modelled, \(\braket{\mathcal{E}[\rho(t)]}_\text{av}=75\%\).

In addition to observing minimal fluctuations in the efficiency, we find that \(\braket{\mathcal{E}[\rho(t)]}_\text{av}\) is closely reproduced (97\% rather than 96\% for the global model) by \(\mathcal{E}[\rho_\text{av}]\), the efficiency of the time-averaged state. Together, these results indicate that, despite short-time fluctuations in the state, the system evolves continuously through states that are similar in character both to each other and to their time average.

\subsection*{An artificial light-harvesting system}
\begin{figure}
\centering
\includegraphics[height=10.4cm]{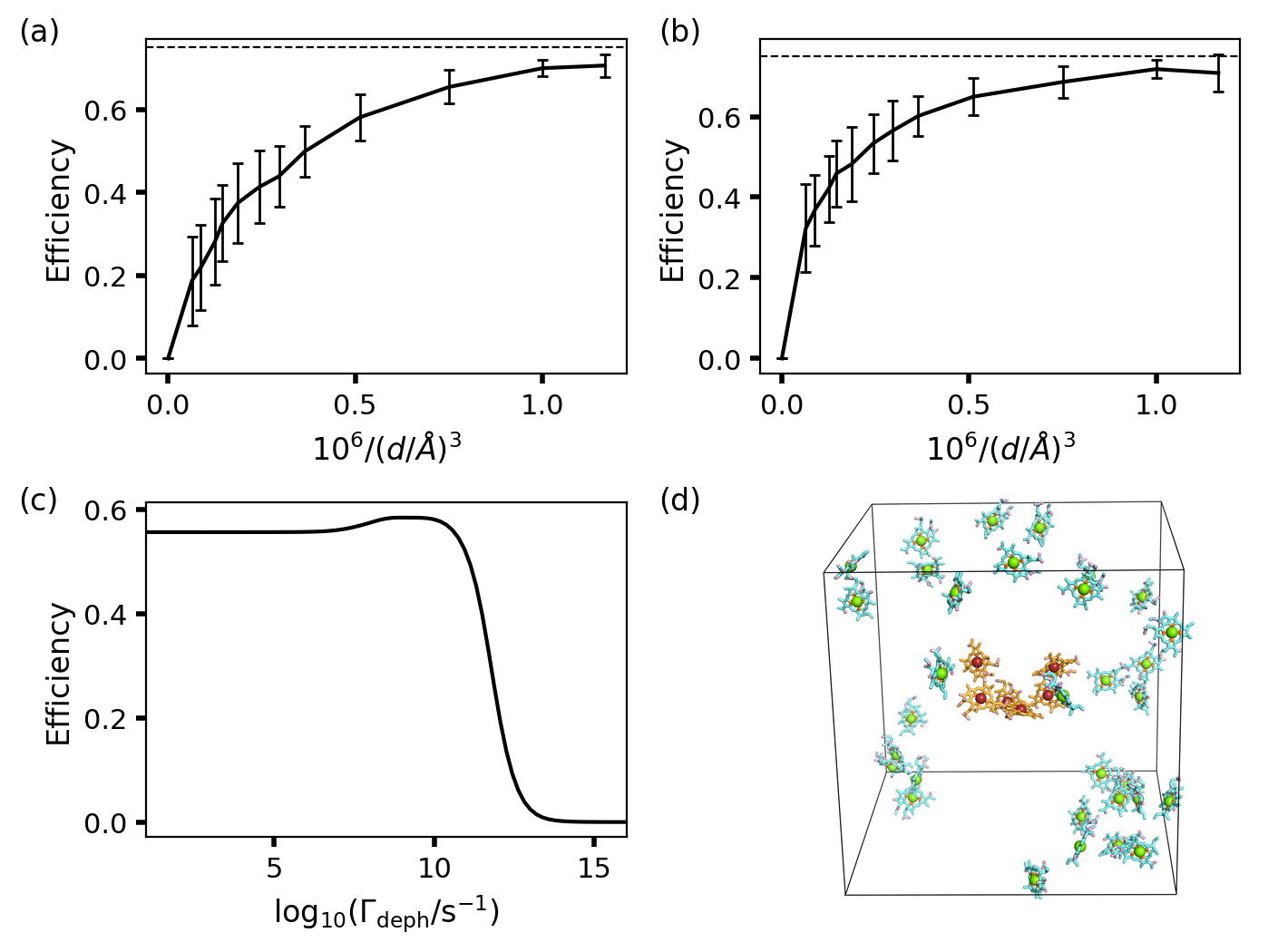}
\caption{Efficiency as a function of chromophore concentration in a (a) cubic ($d\times d\times d$) box or (b) pseudo-2D ($d\times d\times 3 \mbox{\AA}$) slab of 32 chromophores arranged randomly around a centrally placed reaction centre. Each concentration was sampled 32 times. Range bars indicate one standard deviation above and below the mean efficiency. Horizontal dotted lines indicate, for comparison, the efficiency of the LHI-RC complex in purple bacteria, using the local dephasing model. In each sample, the efficiency was averaged over the evolution from 10 ps to 100 ps. (c) Efficiency as a function of dephasing rate for one arrangement of chromophores in a cubic box with $d=125\;\mbox{\AA}$. (d). Example structure of a chromophore box with $d = 95\;\mbox{\AA}$ (corresponding to \(10^6/(d/\mbox{\AA})^3 = 1.17\)).}
\label{fig:liquidefficiency}
\end{figure}

The highly symmetric ring structures of natural photosynthetic antenna systems are often though to imply a special physical significance for the particular spatial arrangement of chromophores (such as maintaining long-lived coherence). We investigated this hypothesis by calculating the efficiency of a box of 32 chromophores (for comparison with LHI) placed and oriented randomly around a central reaction centre. The `antenna' chromophores were not allowed within 3 \AA\  of each other or within 45 \AA\  of the middle of the reaction centre ({\it SI Appendix}). The dimensions of the box were adjusted to alter the concentration of antenna chromophores without changing the ratio of chromophores to reaction centre. Relaxation was implemented using the local dephasing model (\(\mathcal{L}_\text{deph}\)), which is sensitive to chromophore spacing and particularly appropriate to the lower concentrations investigated. Each concentration was sampled 32 times with different random configurations of the chromophores. The results are shown in figure \ref{fig:liquidefficiency}. 

Two clear trends are visible. Firstly, as the concentration of chromophores is reduced, the mean efficiency decreases. More weakly coupled chromophores do not transfer the exciton as quickly to the reaction centre (if at all) so more excited state is lost through non-radiative decay and emission processes.  Secondly, the standard deviation in the efficiency increases with decreased concentration. This can by understood by considering that in a small box, at high concentration, the chromophores will always be close to each other and to the reaction centre.  At lower concentrations, the random placement of chromophores will sometimes result in clusters of closely-spaced, strongly coupled chromophores and other times in all the chromophores being far apart and only weakly coupled. Crucially, when the chromophores were closely spaced, there was very little variation in the efficiency over the 32 sampled configurations, indicating that the orientation of the chromophores was not significant.

The same behaviour was observed whether the chromophores were placed freely in 3D (figure \ref{fig:liquidefficiency}a), as they might be in a liquid phase, or confined to a pseudo-2D slab (figure \ref{fig:liquidefficiency}b), to mimic the layered structure of natural antenna networks.

Whilst chromophore spacing is an important factor for determining efficiency, it is equally important to balance the rates of competing processes in the system. Figure \ref{fig:liquidefficiency}c illustrates the impact of varying the dephasing rate. There is a sharp drop in efficiency when the dephasing rate exceeds \(~100\) ps\(^{-1}\), where continual measurement by the environment prevents the system evolving. Typically, simulations have shown that efficiency drops off at low dephasing rates too, as the exciton is trapped by Anderson localisation. However, in our calculations this effect is largely mitigated by the coupling to slow vibrations.

\section*{Discussion}

We have examined the nature and evolution of the quantum state of a photosynthetic antenna system under constant illumination by a natural, incoherent light source. When talking about the state of light harvesting systems it is common to refer only to the part of the state that contains exactly one exciton.

However, it is worth noting that the overall quantum state is, to a very good approximation, the zero-exciton ground state. This contrasts with the often-employed picture of photon absorption triggering a large change in the excitonic state of a single chromophore. While this picture is relevant for understanding ultrafast spectroscopic experiments, it has little bearing on the phenomenon of photosynthesis in sunlight.

Within the detailed model we have constructed of the many processes governing the state and dynamics of the excitonic subsystem, accurately capturing the effect of interacting with a vibrational environment is particularly challenging, since the similar strengths of inter-system and system-bath coupling places the problem uncomfortably between the ranges of validity of commonly used approximations \cite{Fassioli2013a,Olaya-Castro2011}. The two approaches we have employed embody the two opposing regimes of strongly coupled chromophores interacting collectively with the bath (\(\mathcal{L}_\text{therm}\)) and weakly coupled chromophores interacting independently with a local bath (\(\mathcal{L}_\text{deph}\)). Both methods make the markovian approximation so will not capture the small non-markovian effects that may arise from strongly coupled discrete vibrational modes. Although neither approach can, on its own, reliably describe the effect of the vibrational environment, we obtain valuabe insights by observing features that arise from both descriptions.

Firstly, we conclude that the state of the excitonic subsystem is at all times highly mixed with little or no coherence. Indeed, both descriptions indicate that, on the timescale of the sink (around 10 ps), the average state is close to the maximally mixed within each ring subsystem. In other words, exciton density is distributed equally between equivalent sites in an incoherent manner. Over the whole antenna, the distribution of excited state population is skewed towards LHI because of the energy gradient between successive ring components and the large number of chromophores \cite{Blankenship2002,Hemmig2016}. The importance of this effect is exemplified by the low efficiency obtained using the local dephasing model, which spreads exciton density uniformly over the whole antenna system.

The coupling to a vibrational environment, and the consequent relaxation, is an important mechanism for transporting excitations through the antenna and overcoming disorder-driven localization. The role of dephasing in exciton transport has been noted before and is sometimes referred to as environment-assisted transport \cite{Pelzer2014,Manzano2013,Caruso2009}. What we additionally see here is that slow vibrational modes keep the exciton density moving continuously around the antenna. This is in good agreement with atomistic simulations of LHII, which also show excitons moving perpetually around the system, with a bias towards lower energy ring components \cite{Sisto2017}. By allowing exciton density to explore the whole antenna, the vibrational environment performs the role that was previously postulated to be the purpose of long-lived coherences. It is therefore unsurprising that, along with others \cite{Kassal2013,Manzano2013,Jumper2018}, we observe that excitonic coherence is not a necessary condition for high efficiency.

Without the requirement for coherence, the rigid structural requirements that have been put forward to maintian coherence can be relaxed \cite{Hemmig2016,Olejko2017,Buckhout-White2014}. Indeed, the results of figure \ref{fig:liquidefficiency} demonstrate that the orientation of chromophores has little bearing on the antenna efficiency. Instead, the typical nearest neighbour distance appears to be the key determinant of efficiency. However, effects such as concentration quenching and the energetic cost of synthesising chromophores (which we did not include in our simulation) would make high chromophore concentrations undesirable. We suggest that perhaps the organisation of chromophores into circular structures may simply be a means of avoiding the extremely high pigment concentration that would be required to achieve the right separation in a completely disordered solution. It is possible that the same end could be achieved in an artificial system without the need for complex protein architecture, for example, by anchoring chromophores to a bead.  However, to design an appropriate artificial matrix to support the chromophores, futher investigation would need to be carried out into the role of environment in determining the rates of dephasing and non-radiative decay, which need to be carefully balanced to maintain efficiency.

\begin{acknowledgements}
{We gratefully acknowledge the funding agencies that supported this work: CS was supported through the doctoral training grant of the Engineering and Physical Sciences Research Council (EPSRC); FMWCV was supported through the EPSRC Bristol Centre for Complexity Sciences; SBW is supported by a research fellowship from the Royal Commission for the Exhibition of 1851.
}
\end{acknowledgements}

\bibliography{photo_arxiv.bib}

\end{document}